\documentstyle[prl,aps,multicol,epsfig,amssymb]{revtex}

\begin{document}

\pagestyle{myheadings}
\markboth{Geometry dependent, Natelson {\it et al.}}{Geometry dependent, Natelson {\it et al.}}

\bibliographystyle{prsty}

\draft

\title{Geometry dependent dephasing in small metallic wires}

\author{D. Natelson, R.L. Willett, K.W. West, L.N. Pfeiffer}

\address{Bell Laboratories, Lucent Technologies, Murray Hill, NJ 07974}

\date{\today}

\maketitle

\begin{abstract}

Temperature dependent weak localization is measured in metallic
nanowires in a previously unexplored size regime down to width 
$w=5$~nm.  The dephasing time, $\tau_{\phi}$, shows a low temperature
$T$ dependence close to quasi-1D theoretical expectations 
($\tau_{\phi} \sim T^{-2/3}$) in the narrowest wires, but exhibits
a relative saturation as $T \rightarrow 0$  for wide samples
of the same material, as observed previously.  As only sample
geometry is varied to exhibit both suppression and divergence
of $\tau_{\phi}$, this finding provides a new constraint on
models of dephasing phenomena.

\end{abstract}


\begin{multicols}{2}

Quantum mechanical decoherence in metals is an outstanding problem in
condensed matter physics.  Magnetotransport measurements in a number
of quasi-1D and quasi-2D systems at the smallest accessible size
scales have shown an unexpected saturation in the weak localization
magnetoresistance (WLMR) at low temperatures, interpreted as a
saturation of the coherence time $\tau_{\phi}$ as $T\rightarrow
0$\cite{Mohanty:97,Webb:98}.  These observations run counter to
theoretical expectations \cite{Altshuler:82,LeeRama:85,Imry:98}, since
the inelastic processes that cause decoherence are expected to freeze
out as $T \rightarrow 0$, implying that $\tau_{\phi}$ should diverge
in this limit.  In diffusive, metallic, quasi-1D systems the expected
form of the divergence is $T^{-2/3}$ due to electron-electron
scattering\cite{Altshuler:82}.  Subsequent experiments have shown
reasonable agreement with theory in degenerately doped semiconducting
wires\cite{Gershenson:98}, and a strong material dependence of the
saturation in evaporated metals\cite{Gougam:99}.  This experimental
situation has prompted a number of
theoretical\cite{Mohanty:97:2,Altshuler:98,GZ:98,Maschke:99,Imry:99,Zawadowski:99,Krishnan:99,Wang:00}
responses.  An intrinsic saturation of $\tau_{\phi}$ to a finite value
as $T \rightarrow 0$ would have profound implications for the ground
state of metals and might indicate a fundamental limitation in
controlling quantum coherence in conductors.  The physical mechanism
underlying the saturation of the WLMR remains a subject of
controversy.  Establishing a controlled experimental parameter
influencing this saturation would provide a starting point for
modeling possible dephasing mechanisms.

The WLMR results from pairs of time-reversed loop trajectories that
contribute coherently to the
conductance\cite{Altshuler:82,LeeRama:85,Imry:98}. In a system with a
diffusion constant $D$, the relevant scale for coherence is $L_{\phi}
\equiv \sqrt{D \tau_{\phi}}$, and the scale for thermal smearing is
$L_{T} \equiv \sqrt{\hbar D/k_{\rm B}T}$.  The
quasi-1D regime occurs in samples of width $w$ and thickness $t$ when
$L_{\phi},L_{T} > w,t$, while quasi-2D behavior is expected when
$t<L_{\phi},L_{T}<w$.  Threading magnetic flux through a typical trajectory
suppresses the contribution of such loops, resulting in a MR whose
size and field-scale reflect the sample geometry and
$L_{\phi}$.  Analysis of the WLMR provides an estimate of
$\tau_{\phi}$.

In this Letter, we present transport studies of AuPd nanowires with a
range of widths showing that extremely narrow wires in an unexplored
quasi-1D regime ($w=5$~nm) have a WLMR that agrees with the
theoretical functional $T$ dependence from 4.2~K to below 0.1~K;
increasing sample width in this same material suppresses the
temperature dependence, with quasi-2D wires ($w=1.2~\mu$m) having WLMR
consistent with the saturation seen in previous quasi-2D studies of
this metal\cite{Giordano:87,Webb:98}.  Data at intermediate widths
indicate an evolution from one behavior to the other.  This is the
first observation that the phase coherence as a function of
temperature of a metal can be tuned from suppressed to diverging by
varying an externally controlled parameter, in this case sample
geometry.  Access to this regime is achieved by using metal wires
narrower than 50~nm.

\begin{figure}
\epsfclipon
\epsfxsize=8.5cm
\epsfbox{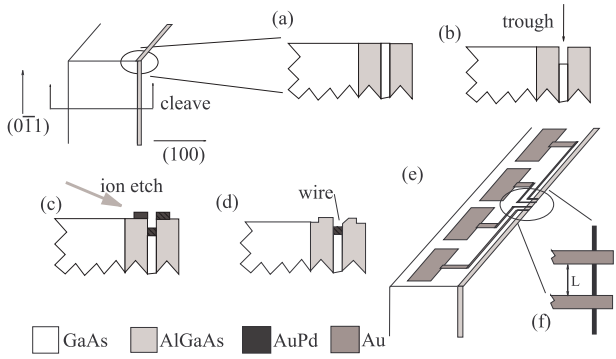}
\caption{The edge wire fabrication process.  (a) MBE-grown substrate
is cleaved; (b) EBL is used to expose the thin GaAs layer, which is
selectively etched to produce a trough; (c) Wire material is
deposited, the resist is lifted off, and a directional ion etch
removes excess material; (d) a nanowire is left in the trough, ready
for further EBL to define leads,(e); Wire length $L$ is defined by
spacing of voltage leads, as shown in (f).}
\label{fig:cartoon}
\end{figure}

Table \ref{tab:samples} summarizes sample dimensions and properties.
The smallest width samples (A-F) were fabricated using a
nonlithographic ``edge'' technique\cite{Natelson:2000a} (see
Fig. \ref{fig:cartoon}), while samples G-K were defined using electron
beam lithography (EBL) and liftoff processing on undoped (100) GaAs
substrates.  The edge technique uses selective chemistry to produce
precise nanoscale relief on the cleaved (0$\bar{1}$1) face of an
undoped MBE-grown GaAs/AlGaAs wafer.  Metal deposition and directional
etching are then used to produce nanowires with widths set by MBE
growth precision.  The wire material for each sample consisted of a
1~nm Ti adhesion layer followed by 7.5~nm of Au$_{0.6}$Pd$_{0.4}$
deposited by electron beam evaporation at a base pressure of 2$\times
10^{-7}$ Torr.  For all samples a lead frame was then defined using
EBL, and made by e-beam evaporation of 2.5~nm of Ti and 90~nm of Au
with subsequent liftoff.

Sheet resistances of codeposited films were measured at 4.2~K.
Typical resistivity of deposited films is found to be
$\sim$24~$\mu\Omega$-cm, consistent with other investigators'
evaporated AuPd films\cite{Giordano:87}.  Free electron values for the
density of states $\nu$ and Fermi velocity $v_{\rm F}$ for pure
Au\cite{Ashcroft} were combined with the Einstein relation to estimate
the diffusion constant $D$ and the elastic mean free path $\ell \sim
2-3$~nm from measured film resistivities.

Contact resistances between the leads and wires were on the order of a
few Ohms, while typical lead resistances were 100-200$\Omega$.  At the
lead-wire contact point, the leads were $\sim 0.5\mu$m wide.  The
interlead spacing $L$ is the distance between ``inner'' edges of
voltage leads (see Fig. \ref{fig:cartoon}).  These lengths were
considerably greater than $\ell$ and $L_{\phi}$ (as analyzed below),
suppressing ``non-local'' four-terminal resistance
effects\cite{WashWebb}.

Samples were mounted on a dilution refrigerator, with ``bath''
temperature measured using a calibrated Ge resistance thermometer.
Resistances were measured using lock-in techniques at a frequencies
$\le$ 17~Hz.  For samples A-E, two- and four-terminal measurements
were employed with similar results, while the remaining samples were
only examined with a four-terminal bridge method.  Measurement
currents were maintained at levels low enough to avoid Joule heating
(0.05~nA for samples A-E, 0.5~nA for sample F, $<$5~nA for samples
G-K).  Two checks on this were performed: $R$ continued to vary down
to the lowest temperatures, and there both $R(H=0)$ and $R(H)$ were
unchanged when the measuring current was reduced by a factor of two.
Magnetic fields as large as 8~T were applied transverse to the
direction of current in the wires.

\begin{figure}
\epsfclipon
\epsfxsize=8.5cm
\epsfbox{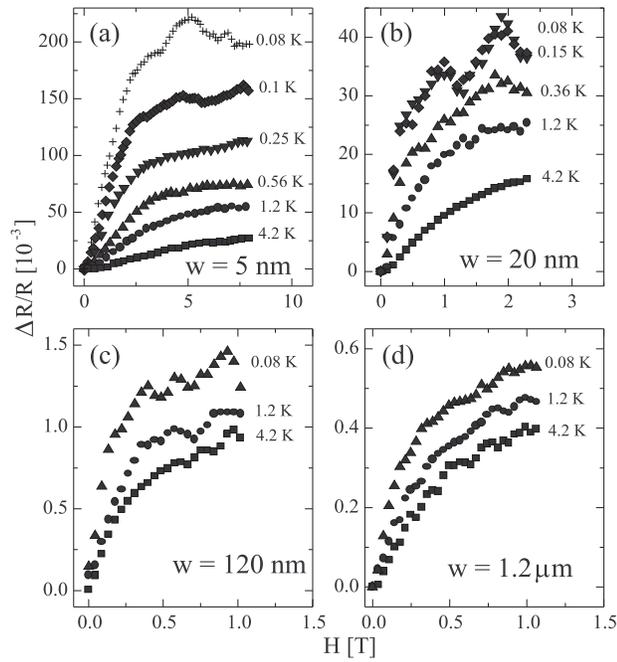}
\caption{Perpendicular magnetoresistance curves for 
representative samples, plotted as $(R(T,H)-R(T,0))/R(T,0)$ vs. $H$.
The WLMR is symmetric about zero field. (a) Sample A, $H$ along
(100),$w = 5$~nm, quasi-1D; (b) Sample F, $H$ along (0$\bar{1}$1), $w
= 20$~nm, quasi-1D; (c) Sample G, array of 19 wires in parallel, $H$
along (100), $w = 0.12~\mu$m, intermediate dimensionality; (d) Sample
K, $H$ along (100), $w = 1.2~\mu$m, quasi-2D.  }
\label{fig:magres}
\end{figure}

Figure \ref{fig:magres} shows magnetoresistance data for various
sample widths and summarizes our key finding: varying sample width
alters the MR temperature dependence.  The positive sign of the MR is
consistent with previous observations \cite{Giordano:87} that AuPd is
a strong spin-orbit scattering system.  The data are symmetric in
field, so only positive field direction sweeps are shown.  The data
for the 5~nm-wide sample (sample A) are shown in
Fig. \ref{fig:magres}a; MR data for $H$ in the other transverse
direction were similar.  The shape of the MR curves is consistent with
a quasi-1D dimensionality ($L_{\phi} > d,t$).  The MR is
temperature-dependent over the entire range, increasing from
$\sim$2.5\% at 4.2~K to roughly 15\% at 100~mK, and the field scale at
which $\Delta R/R$ saturates moves to smaller fields as $T$ is
decreased.  At temperatures less than 0.5~K aperiodic variations of
$\Delta R/R$ with $H$ are visible.  Such sample-specific fluctuations
varied from cooldown to cooldown, and are universal conductance
fluctuations as a function of external field.

Compare these curves with the analogous MR data for a quasi-2D
1.2~$\mu$m-wide wire (sample K) shown in Fig. \ref{fig:magres}d.  The
size and field scale of the MR effect are consistent with the
dimensionality and resistance per square of the sample.  However, the
MR is only weakly dependent on temperature over the entire range,
varying from 0.04\% to 0.055\%, in contrast to the narrow wire data.
Fig. \ref{fig:magres}c shows MR traces for an array of
0.12~$\mu$m-wide wires, expected to be of intermediate dimensionality
(almost quasi-2D at 4.2~K, almost quasi-1D at 0.1~K).  For this
intermediate size, the temperature dependence of the overall MR effect
is slightly larger in size than the results for the widest wires.
Fig. \ref{fig:magres}b shows MR data for 20~nm-wide wires, expected to
be in the quasi-1D limit over the entire temperature range; the
temperature dependence of MR data with $H$ in the other transverse
direction was similar.  The MR temperature dependence is more
pronounced than the 0.12~$\mu$m case, but not as large as in the 5~nm
wires.  The raw data in Fig. \ref{fig:magres} show that the
temperature dependence of the MR in a single material increases
substantially as the width of the wires is reduced, independent of the
analysis of the MR as a measure of electronic coherence in these
samples.

We use the theory of weak localization to extract a characteristic
dephasing time, $\tau_{\phi}$, from the MR data.  Analytic MR
predictions exist for samples of definite dimensionality, and an
internally consistent analysis requires, {\it e.g.}, the extracted
$\sqrt{D \tau_{\phi}} > d,t$ if a quasi-1D formula was used.  The
relevant MR predicted\cite{Altshuler:82,Echternach:93} for the
quasi-1D case is:
\begin{equation}
\frac{\Delta R}{R} = \frac{1}{\pi \sqrt{2}} \frac{e^2}{\hbar}
\frac{R}{L} \sqrt{D \tau_{\rm N}} \times
f \left( \frac{2 \tau_{\rm N}}{\tau_{\phi 0}} \right),
\label{eq:1dwlmagres}
\end{equation}
with $f(x) \equiv {\rm Ai}(x)/{\rm Ai'}(x)$, where Ai$(x)$ is the Airy
function.  Here $\tau_{\rm N}$ is the time associated with coherence
loss due to small angle Nyquist electron-electron scattering.  The
rate $\tau_{\phi0}^{-1}$is the sum of large energy ($\sim k_{\rm B}T$)
scattering processes at $H=0$ and the magnetic contribution,
$\tau_{H}^{-1} = D W^{2}/ 12 L_{H}^{4}$, with $L_{H} =
\sqrt{\hbar/2eH}$.  Here $W$ is the sample dimension transverse to the
applied magnetic field.  To minimize the number of fitting parameters,
we fix $W$ at 5~nm for samples A-E and 20 nm for sample F; we use $D$
as inferred from the codeposited film; and we set the time
$\tau_{\phi0}(H=0)$ to some long value (100~ns).  Using measured
values of $R$ and $L$ at each temperature, the only remaining
variational parameter is $\tau_{\rm N}$.  We limit the fitting with
Eq. (\ref{eq:1dwlmagres}) to the field regime $L_{H} > W$ and avoid
large conductance fluctuations.  The theoretical prediction for the
quasi-1D Nyquist scattering time is given by:
\begin{equation}
\tau_{\rm N,th} = \left( \frac{\hbar^{2} L}{e^{2} \sqrt{2D} k_{\rm B}T R}
\right)^{2/3}.
\label{eq:tauN}
\end{equation}

For the quasi-2D case (samples H-K), the appropriate magnetoresistance
prediction is\cite{Hikami:80,Giordano:87}:
\begin{equation}
\frac{\Delta R}{R} = \frac{R_{\Box} e^2}{4 \pi^2 \hbar} \left[ \psi
\left( \frac{1}{2} + \frac{L_{H}^{2}}{2  D \tau_{\phi}} \right) - \ln
\left( \frac{L_{H}^{2}}{2 D \tau_{\phi}} \right) \right],
\label{eq:2dwlmagres}
\end{equation}
where $\psi(x)$ is the digamma function.  To fit the data we use the
measured resistance per square $R_{\Box}$, $D$ as inferred from the
resistivity, and allow $\tau_{\phi}$ in Eq. (\ref{eq:2dwlmagres}) to
vary.

\begin{figure}
\epsfclipon
\epsfxsize=8.5cm
\epsfbox{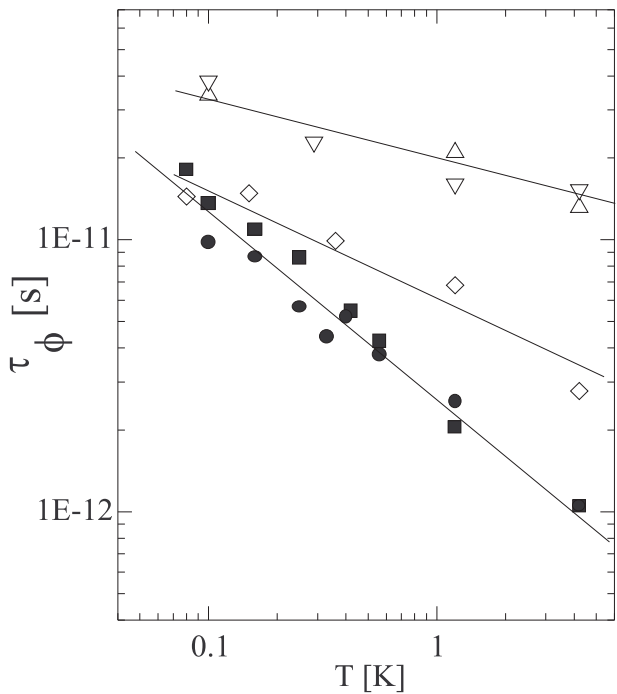}
\caption{Coherence times as a function of temperature, extracted using
Eqs. (1) and (3) for samples (A =
$\blacksquare$, B=$\bullet$,F=$\diamond$) and
(J=$\bigtriangledown$, K=$\bigtriangleup$), respectively.  Data for (J,K)
have been vertically offset (multiplied by 2.5) for clarity; lines are
a guide to the eye, and top to bottom correspond to power laws of $T^{-0.22}$, $T^{-0.40}$, and $T^{-0.67}$.}
\label{fig:tauphi}
\end{figure}

Figure \ref{fig:tauphi} shows the results of fitting the MR data for
representative samples, 5~nm-wide samples A and B, 20~nm-wide sample
F, and wider samples J and K.  The other samples show similar
behavior.  The scatter in the fit parameter $\tau_{\rm N}$ is
dominated by variation in the MR data due to universal conductance
fluctuations\cite{WashWebb,Feng:91,Natelson:2000a}.  For samples A-E,
$\tau_{\rm N}(T)$ was fit to the form $K T^{p}$, yielding $p = -0.59
\pm 0.05$, consistent with the predicted $T$ dependence of
Eq. (\ref{eq:tauN}).  The prefactor $K$ found in samples A-E
is approximately a factor of four smaller than that predicted by
Eq. (\ref{eq:tauN}), consistent with previous
investigations\cite{Echternach:93,Gershenson:99,Gougam:99}.

In contrast with this consistency between the experimental results and
theory, consider the $\tau_{\phi}(T)$ parameters extracted from wide
samples J and K using Eq. (\ref{eq:2dwlmagres}).  The weak temperature
dependence of $\tau_{\phi}$ in these wider samples is as seen in
previous studies of quasi-2D samples of AuPd\cite{Giordano:87}.  Note
that the quasi-2D prediction for Nyquist dephasing is an even steeper
temperature dependence than the quasi-1d case, $\tau_{\phi,{\rm 2D}}
\sim T^{-1}$\cite{Altshuler:82}, and for $R_{\Box}\approx 32\Omega$,
$\tau_{\phi,{\rm 2D}} \sim 6\times 10^{-9}$~s at 1~K.  These
experimental findings support an empirical picture in this material of
$\tau_{\phi}$ evolving from a form consistent with e-e interactions to
one not understood as sample width is increased toward 2D.  Attempts
to analyze the suppressed MR data using a sum of two different
dephasing rates are complicated by the small size of the MR effects in
the widest samples.

Proposed mechanisms for the anomalous saturation of $\tau_{\phi}$ as
$T \rightarrow 0$ must be considered in light of the size dependence
reported here.  The cause of the saturation in the widest wires
remains unknown at this time, and may well involve an additional
phase-breaking mechanism with a weak $T$ dependence over the interval
examined.  Any explanation based on the intrinsic properties of the
material must reconcile the observed variation of the saturation with
sample size, and with metal, since $\tau_{\phi}$ in Ag wires has been
seen to not saturate down to 40~mK, while saturation was seen at
$\sim$ 700~mK in Cu wires with similar parameters\cite{Gougam:99}.  
Careful examination of these metals over size ranges is warranted.

Our observations also constrain possible extrinsic sources of
dephasing.  Since overall sample resistances for the widest and
narrowest wires were of the same order ($\sim 10~{\rm k}\Omega$), it
is unlikely that external RF noise\cite{Gershenson:98,Altshuler:98}
can account for the difference in WLMR behaviors.  

One plausible
explanation for the geometry-dependence of the MR is competition
between Nyquist scattering and an unknown dephasing mechanism.  Both
processes presumably exist in all the samples, but for fixed disorder
smaller sample dimensions enhance the Nyquist process (see
Eq. (\ref{eq:tauN})) by increasing $R/L$.  It is possible that in
sufficiently narrow wires, the Nyquist dephasing rate $\tau_{\rm
N}^{-1}$ becomes more rapid than the competing process, while in wide
wires the other process could dominate as the Nyquist scattering rate
is reduced.  Alternately, the unknown dephasing mechanism may be
suppressed as sample size is reduced below some crucial lengthscale.
Detailed studies of the size dependence in this and other materials
and extensions to lower temperatures, while significant experimental
challenges, may help distinguish these possibilities.

We have examined magnetotransport as $T\rightarrow 0$ in AuPd
nanowires down to wire widths substantially below 10~nm, demonstrating
that a single material system can exhibit either saturating or
diverging magnetoresistance behavior depending on sample geometry.
When analyzed within the framework of WL theory, this translates into
a difference in inferred dephasing behaviors.  The narrowest wires
seem qualitatively consistent with the predictions for Nyquist
dephasing in quasi-1D systems, while wide wires exhibit a saturation
of $\tau_{\phi}$ similar to that seen in previous investigations.
With the evidence that geometry can tune MR behavior between
saturating and nonsaturating regimes, we have a new tool for examining
the properties of $\tau_{\phi}$.

We thank B.L. Altshuler, K. Baldwin, R. dePicciotto, E.M.Q. Jariwala,
P. Mohanty, P.M. Platzman, S. Simon, and C.M. Varma for valuable
discussions.

\end{multicols}


\begin{table}
\begin{tabular}{c|c|c|c|c|c}
\hline \hline
Sample & $w$ & $t$ & $L$ & $D$ & $R/L$ or $R_{\Box}$ \\
 & [nm] & [nm] & [$\mu$m] & [m$^{2}$/s] & [@ 4.2~K] \\
\hline
A & 5 & 7.5 & 1.5 & 1.2 $\times 10^{-3}$ & 19.2 k$\Omega$/$\mu$m \\
B & 5 & 7.5 & 1.5 & 1.2 $\times 10^{-3}$ & 17.0 k$\Omega$/$\mu$m \\
C & 5 & 7.5 & 1.5 & 1.5 $\times 10^{-3}$ & 14.9 k$\Omega$/$\mu$m \\
D & 5 & 7.5 & 1.5 & 1.5 $\times 10^{-3}$ & 14.1 k$\Omega$/$\mu$m \\
E & 5 & 7.5 & 0.75 & 1.5 $\times 10^{-3}$ & 14.8 k$\Omega$/$\mu$m \\
F & 20 & 7.5 & 1.5 & 1.5 $\times 10^{-3}$ & 3.6 k$\Omega$/$\mu$m \\
G & 120 & 7.5 & 6.6 & 1.5 $\times 10^{-3}$ & 265 $\Omega$/$\mu$m \\
H & 1100 & 7.5 & 380 & 1.5 $\times 10^{-3}$ & 31.5 $\Omega$/$\Box$ \\
I & 1100 & 7.5 & 380 & 1.5 $\times 10^{-3}$ & 31.5 $\Omega$/$\Box$ \\
J & 1250 & 7.5 & 380 & 1.5 $\times 10^{-3}$ & 32.3 $\Omega$/$\Box$ \\
K & 1250 & 7.5 & 380 & 1.5 $\times 10^{-3}$ & 32.3 $\Omega$/$\Box$ \\
\hline
\end{tabular}
\caption{Samples used in magnetotransport measurements.  To minimize
``magnetofingerprint'' effects, Sample F is an average of two
1.5~$\mu$m segments, while Sample G is an array of 19 wires in
parallel, for which single-wire parameters are listed.}
\label{tab:samples}
\end{table}


\end{document}